\documentstyle[12pt]{article}
\title{Diffeomorphism algebra of two dimensional free massless scalar field with signature change}
\author{F. Darabi$^{a,b}$, M. A. Jafarizadeh$^{a,c,}$    \thanks{E-mail: tabriz\_u@vax.ipm.ac.ir}
and A. Rezaei-Aghdam$^a$ \vspace{10mm}\\ 
$^a$ {\small Faculty of Physics, Tabriz University, Tabriz 51664, Iran.} \\ 
$^b$ {\small Department of Physics, Shahid Beheshti University, Tehran 19834, Iran.} \\
$^c$ {\small Institute for Studies in Theoretical Physics and Mathematics, Tehran, 19395-1795, Iran.}}
\begin{document}
\maketitle
\vspace{15mm}
\begin{abstract}
We study a model of free massless scalar fields on a two dimensional cylinder 
with metric that admits a change of signature between Lorentzian and Euclidean 
type (ET), across the two timelike hypersurfaces (with respect to Lorentzian 
region). Considering a long strip-shaped region of the cylinder, denoted by an 
angle $\theta$, as the signature changed region it is shown that the energy 
spectrum depends on the angle $\theta$ and in a sense differs from ordinary one for low 
energies. Moreover diffeomorphism algebra of corresponding infinite conserved 
charges is different from ``virasoso'' algebra and approaches to it at higher energies.
The central term is also modified but does not approach to the ordinary one at higher energies.
\end{abstract}

\section{Introduction}
The initial idea of signature change is due to Hartle, Hawking and Sakharov \cite{HHS}
which makes it possible to have a spacetime with Euclidean and Lorentzian 
regions in quantum gravity. It has been shown that the signature change may happen even in classical general relativity \cite{CSC}. There are 
two different approaches to this problem: continuous and discontinuous 
signature changes. In the continuous approach, in going through Euclidean 
region to Lorentzian one the signature of metric changes continuously. 
Hence the metric becomes degenerate at the border of these regions.
But in the discontinuous approach the metric is nondegenerate everywhere and  
discontinuous at the border of Euclidean and Lorentzian regions. 
In quest for effect of signature change in physical problems, mostly its 
effect on the propagation of massless boson has been rather deeply studied \cite{D}. 
Dray and $et al$ have shown that the phenomena of particle production 
can happen for propagation of scalar particle in spacetime with heteric signature. 
They have also obtained a rule for propagation of massless scalar field on 
a two dimensional space-time with signature change \cite{D}.
The effect of signature change on the propagation of plane waves in going 
from Lorentzian to Kleinian region has been studied by Alty \cite{A}. 
They have shown that this kind of signature change leads to unlimited energy 
extraction from Kleinian region. 
There are of course rather a bit of works about the subject of signature 
change, which we do not follow them here. In this paper we follow the 
idea similar to Kleinian type signature change \cite{A} in which it is a spacelike 
coordinate whose associated metric component changes sign discontinuously. 
We study the effect of such signature change over $M=R \times S^1$ manifold 
on the free massless scalar field (boson field), where $R$ represents 
timelike coordinate $\tau$ and the change of signature is induced from the 
change of sign in the metric component associated to the spacelike coordinate 
$\sigma$ on the circle $S^1$.
The model is similar to that of Dray and et al \cite{D} with the a difference 
that the role of space and time are changed. An idea about the topology 
change of the manifold $M=R \times S^1$ over the spacelike hypersurface (circle) 
\cite{Pr} has been one of our motivations in studying the effects that such a 
signature change produces.
The standard investigation of ordinary free massless scalar field on the 
manifold $M=R \times S^1$ with a pure Lorentzian signature gives rise to 
discrete energy spectrum of integer values and an infinite number of conserved 
charges called ``virasoro generators'' \cite{B}. These generators form a diffeomorphism 
algebra called ``virasoro algebra'' which gets a central term after quantization \cite{B}. 
Introducing a region of Euclidean type (ET) signature, denoted by an angle 
$\theta$ as a long strip-shaped region of the cylinder $M=R \times S^1$ 
enclosed by a Lorentzian region, affects these standard results such that 
the energy spectrum and diffeomorphism algebra are modified; specially the 
corresponding central term is a complicate function of energy and the angle $\theta$. The modifications 
of spectrum and diffeomorphism algebra tends to disappear at higher values of energy $\omega$ 
but the central term does not approach to the standard central term at higher energies.
Thus it appears to be worth investigating this model.
The paper is organized as follows:
\newline 
In section ${\bf 2}$, using the notation of reference \cite{D}, the signature change over 
$S^1$ has been throughoutly studied, where by requiring the continuity of currents 
accross the border of (ET) and Lorentzian regions we reach a junction 
condition. In section ${\bf 3}$, the equation of motion for massless scalar boson has 
been solved in both regions. Then imposing the appropriate junction conditions,
a quantization condition is imposed on the values of spectrum, $\omega$, and 
real distributional solutions have been constructed. The section is ended by 
investigating their completeness and orthogonality. In section ${\bf 4}$, using 
the Poisson bracket of real scalar field with its conjugate momentum in both 
regions, the Poisson bracket of normal modes $\alpha_\omega$ has been obtained.
In section ${\bf 5}$, using the conservation of energy-momentum tensor distributions 
on both sides of the border of signature change in light cone coordinates,
infinite number of conserved quantities $L_\omega$ have been obtained.  
In section ${\bf 6}$, the diffeomorphism algebra of $L_\omega$ is constructed. 
In section ${\bf 7}$, we expand $L_\omega$ in normal modes. Finally in section ${\bf 8}$, we      
quantize this model by Dirac canonical quantizition method and show that the 
diffeomorphism algebra gets a central term which has been calculated for higher
energies. The paper ends with a conclusion and an appendix explaining the derivation 
of structure constants of the algebra.
\section{Definition of the model}
In this section we use the symbols of Ref.\cite{D} .
We take the lagrangian of free massless scalar field $\phi$ with signature change in two 
dimensions as
\begin{equation}
{\cal L}\star1=d\phi\wedge \star \: d\phi
\end{equation}
where $d$ is the exterior derivative and $\star$ is the Hodge star given by 
\begin{equation}
\star1{\mid_{U^\pm}}=\epsilon^\pm \sqrt{\mid g \mid}d\tau \wedge d\sigma \mid_{U^\pm}
\end{equation}
where $\epsilon^\pm$ takes the values $\pm 1$ according to the orientation of 
the coordinates $\tau$ and $\sigma$ in both regions $U^\pm$ of different signatures.
Morever we assume that the free massless scalar field propagates on a 2-dimensional 
manifold $M=R \times S^1$ (the circle $S^1$ represents ``space'' and the real line 
represents ``time'') with the following metric
\begin{equation}
dS^2= -\: d\tau^2 + g (\sigma)\: d \sigma^2
\end{equation}
where $\tau$ is timelike coordinate and $\sigma$ is a kind of periodic spacelike coordinate 
with period $2\pi$, and $g(\sigma)$ is a periodic function of 
$\sigma$, which takes +1 for Lorentzian region and -1 for (ET) one.
Let us introduce an angle $\theta$ as a segment of the coordinate $\sigma$ for which
$g(\sigma)$ is given by
\begin{equation}
g(\sigma)=\left \{ \begin{array}{ll}
-1 &\hspace{5mm} 0<\sigma<\theta + Mod\:  2\pi \hspace{7mm}\sigma\in U^+ \\
+1 &\hspace{5mm} \theta<\sigma<2\pi + Mod\:  2\pi\hspace{5mm}\sigma\in U^-
\end{array}\right. 
\end{equation}
The situation is shown in Fig.1 in which the shaded region has (ET) metric 
and a Lorentzian metric is governed elsewhere.\\
Obviously for every killing vector $X$, we have a conserved current as
\begin{equation}
J_X=i_X d\phi \wedge \star \: d\phi + d\phi \wedge i_X \star d\phi
\end{equation}
where $i_X$ denotes the contraction with $X$.  
Now, using the definition of Hodge star in (2), the currents $J^\pm$ can be written as
\begin{equation}
J^\pm = \epsilon^\pm[\frac{(\partial_\sigma \phi)^2}{- g} -(\partial_\tau \phi)^2]{\sqrt{\mid g \mid}} d\sigma - 2\epsilon^\pm \frac{{\partial_\tau}\phi{\partial_\sigma}\phi}{\sqrt{\mid g \mid}} sgn(g) d\tau               
\end{equation}
where $sgn(g)$ denotes the definition (4).
In order the closed one-form $J_X$ 
to be defined consistently on the manifold $M$, the pullbacks of $J^\pm$ 
to each hypersurface of signature change $\Sigma$ , $\Sigma^\prime$ must 
agree, using of stokes' theorem \cite{D}.
The agreement of pullbacks $-2\epsilon^\pm \partial_\tau\:\phi\: \partial_{\Sigma}\:\phi\: sgn(g)\: d\tau$ 
at each hypersurface $\Sigma$ , $\Sigma^\prime$ together with the assumption 
of continuity of $\phi$ across these hypersurfaces leads to the following 
``junction condition'' at each $\Sigma$ and $\Sigma^\prime$ 
\begin{equation}
\partial_{\sigma_E} \phi\mid_{\Sigma, \Sigma'} = - \: \epsilon \; \partial_{\sigma_L} \phi\mid_{\Sigma, \Sigma'} \hspace{10mm} \epsilon=\frac{\epsilon^+}{\epsilon^-}
\end{equation}                 
Of course the junction condition (7) is similar to the one in \cite{D} 
with the difference that the roles of space and time are interchanged.

\section{Solution of wave equations }
The wave equations obtained from the variation of Lagrangian (1) in (ET) 
and Lorentzian regions are
\begin{equation}
\begin{array}{ll}
(\:{\partial_\tau}^2 + {\partial_{\sigma}}^2\:)\:\phi^E_\omega(\sigma,\tau)=0\\
\\
(\:{\partial_\tau}^2 - {\partial_{\sigma}}^2\:)\:\phi^L_\omega(\sigma,\tau) =0
\end{array}
\end{equation}
where $\phi^E_\omega(\sigma,\tau)$ and $\phi^L_\omega(\sigma,\tau)$ are solutions 
in (ET) and Lorentzian regions respectively.
Obviously these equations can be solved exactly and their solutions are
\begin{equation}
\begin{array}{ll}
\phi^E_\omega(\sigma,\tau) = A_\omega exp\:(-i\omega(\tau + i\sigma)) + B_\omega exp\:(-i\omega(\tau - i\sigma))\\
\\
\phi^L_\omega(\sigma,\tau) = C_\omega exp\:(-i\omega(\tau + \sigma)) + D_\omega exp\:(-i\omega(\tau - \sigma))
\end{array}
\end{equation}
where the coefficients $A_\omega,B_\omega,C_\omega$ and $D_\omega$, as well as the 
energy $\omega$ can be determined by using the junction condition (7) over 
the solutions (9). Assuming $\epsilon^+ =+1$ and $\epsilon^- =+1$ 
for (ET) and Lorentzian regions respectively; the continuity of $\phi_\omega$
and the junction condition (7) are written as 
\begin{equation}
\begin{array}{cc}
\phi^E_\omega \mid_0 = \phi^L_\omega \mid_{2\pi} \hspace{10mm} \phi^E_\omega \mid_\theta =\phi^L_\omega \mid_\theta \\
\\
\partial_{\sigma} \phi^E_\omega \mid_0 = - \partial_{\sigma}\phi^L_\omega \mid_{2\pi}  \hspace{10mm} \partial_{\sigma} \phi^E_\omega \mid_\theta = - \partial_{\sigma} \phi^L_\omega \mid_\theta
\end{array}
\end{equation}
Here we have used the fact that these conditions are satisfied at all times 
along the coordinate $\tau$.
Now, these conditions have  nontrivial solutions for coefficients 
$A_\omega,B_\omega,C_\omega$ and $D_\omega$, only if $\omega$ satisfies the 
following ``quantization condition''
\begin{equation}
cosh \omega \theta\: cos \omega (\theta - 2\pi) =1
\end{equation}
For a given root of the equation (11), the coefficients $A_\omega,B_\omega,
C_\omega$ and $D_\omega$ are
\begin{equation}
\begin{array}{ll}
A_\omega =\frac{1-i}{2}[((exp\:(i\omega(\theta - 2\pi)) -cosh \omega \theta)/(sinh \omega \theta)) + 1 ]  exp\:(2\pi i \omega) D_\omega = a_\omega D_\omega\\
\\
B_\omega =\frac{1+i}{2}[-((exp\:(i\omega(\theta - 2\pi)) -cosh \omega \theta)/(sinh \omega \theta)) + 1 ]  exp\:(2\pi i \omega) D_\omega = b_\omega D_\omega\\
\\
C_\omega =\frac{-i (exp\:( i \omega \theta) - exp\:(2\pi i\omega) cosh \omega \theta)}{exp\:(-2\pi i \omega) sinh \omega \theta} D_\omega = c_\omega D_\omega\\
\\
D_\omega = arbitrary
\end{array}
\end{equation}
where, the arbitrariness of $D_\omega$ makes it possible to fix other 
coefficients $A_\omega,B_\omega$ and $C_\omega$.
The spectrum of this model, $\omega$, is real and $\theta$-dependent. 
It differs from ordinary spectrum (with pure Lorentzian signature) at low 
energies and, as is seen in Fig.2, at high energies it coincides with the 
roots of $cos{\omega(2 \pi-\theta)}$; therefore it,s ``sum over energies''
approaches to ``sum over integers'' at higher energies.\\
Now, we come to the orthogonality of our solutions for different values of 
$\omega$. 
Let us consider two solutions $\phi^E_\omega(\sigma, \tau)$ and 
$\phi^E_{\omega^\prime}(\sigma, \tau)$, corresponding to two different roots of 
equation (11).
Obviously they satisfy the following equations in the (ET) region
$$
\hspace{24mm}\{\partial^2_{\sigma} \phi^E_\omega(\sigma) - \omega^2 \phi^E_\omega(\sigma)\} exp(-i \omega \tau) = 0  \hspace {63mm} (13 - a)
$$
$$
\hspace{24mm}\{\partial^2_{\sigma} \phi^E_{\omega^\prime}(\sigma) -\omega^{\prime^2} \phi^E_{\omega^\prime}(\sigma)\} exp(-i \omega \tau) = 0  \hspace {65mm} (13 - b)
$$
where $ \phi^E_\omega(\sigma) $ and $ \phi^L_\omega(\sigma) $ are $ \sigma $ dependent 
separable solutions.  
Multiplying the first equation by $-\phi^E_{\omega^\prime}(\sigma)$ 
and the second one by $\phi^E_{\omega}(\sigma)$, integrating 
them in (ET) region from zero to $\theta$ and finally adding them up, we get  
\setcounter{equation}{13}
\begin{equation}
(\omega^2 - {\omega^\prime}^2) \int _0^\theta\!\phi^E_\omega(\sigma)\:\phi^E_{\omega^\prime}(\sigma)\,d\sigma - \int _0^\theta\!\partial_{\sigma}(\phi^E_{\omega^\prime}(\sigma)\:\partial_{\sigma}\phi^E_\omega(\sigma) - \phi^E_\omega(\sigma)\:\partial_
{\sigma}\phi^E_{\omega^\prime}(\sigma))\,d\sigma = 0
\end{equation} 
note that we have ignored the $\tau$ integration, since it does not affect 
the result (14).
Similarly, the solutions $\phi_\omega^L(\sigma)$ and $\phi_{\omega'}^L(\sigma)$ 
satisfy the following equations in the Lorentzian region
$$
\hspace{24mm}\partial^2_{\sigma} \phi^L_\omega(\sigma) + \omega^2 \phi^L_\omega(\sigma) = 0  \hspace{88mm} (15-a)     
$$
$$
\hspace{23mm}\partial^2_{\sigma} \phi^L_{\omega^\prime}(\sigma) + \omega^{\prime^2} \phi^L_{\omega^\prime}(\sigma) = 0  \hspace{87mm} (15-b)     
$$
By a similar procedure we will have 
\setcounter{equation}{15}
\begin{equation}
(\omega^2 - {\omega^\prime}^2) \int _\theta^{2\pi}\!\phi^L_\omega(\sigma)\:\phi^L_{\omega^\prime}(\sigma)\,d\sigma + \int _\theta^{2\pi}\!\partial_{\sigma}(\phi^L_{\omega^\prime}(\sigma)\:\partial_{\sigma}\phi^L_\omega(\sigma) - \phi^L_\omega(\sigma)\:
\partial_{\sigma}\phi^L_{\omega^\prime}(\sigma))\,d\sigma = 0
\end{equation}
for the Lorentzian region. Adding equations (14) and (16), and then imposing 
the junction conditions (10) we get
\begin{equation}
(\omega^2 - {\omega^\prime}^2) (\int _0^\theta\!\phi^E_\omega(\sigma)\:\phi^E_{\omega^\prime}(\sigma)\,d\sigma + \int _\theta^{2\pi}\!\phi^L_\omega(\sigma)\:\phi^L_{\omega^\prime}(\sigma)\,d\sigma) = 0
\end{equation} 
Now, if we introduce the distribution \cite{D}
\begin{equation}
\phi_\omega(\sigma)=\Theta^+\:\phi_\omega^E + \Theta^-\:\phi_\omega^L
\end{equation}
where $\Theta^\pm$ are the Heaviside distribution with support in $U^\pm$, 
then we get the following orthogonality condition on the $\tau = const$ hypersurface
\begin{equation}
\frac{1}{2 \pi} \int _0^{2\pi} \! \phi_{\omega} (\sigma)  \phi_{\omega^\prime} (\sigma)  \star 1 =(\delta_{\omega ,\omega^\prime} + \delta_{\omega , -\omega^\prime})<\phi_\omega,\phi_{\omega^\prime}> 
\end{equation}
where the factor $1/2\pi$ is introduced for convenience and the symbol $<,>$
denotes the inner product of solutions. Choosing $D_\omega=1/b_\omega$ 
and substituting $A_\omega,B_\omega,C_\omega$ given by equation (12) in (9), we get 
\begin{equation}
\begin{array}{ll}
\phi^E_\omega(\sigma,\tau)= [(a/b)_\omega exp(\omega \sigma) + exp(-\omega\sigma)] exp(-i\omega\tau)\\
\\
\phi^L_\omega(\sigma,\tau)= [(c/b)_\omega exp(-i\omega\sigma) + (1/b)_\omega exp(i\omega\sigma)]exp(-i\omega\tau)
\end{array}
\end{equation}
\begin{equation}
\begin{array}{ll}
\phi^E_0 = \:[(a/b)_0 +1]\\
\\
\phi^L_0 = \:[(c/b)_0 + (1/b)_0]
\end{array}
\end{equation}
where $(a/b)_\omega$,$(c/b)_\omega$ and $(1/b)_\omega$ are given by
\begin{equation}
\begin{array}{ll}
(a/b)_\omega =  \frac{sin\omega(\theta-2\pi)}{cosh\omega\theta + sinh\omega\theta - cos\omega(\theta-2\pi)} \\
\\
(c/b)_\omega =\frac{(1+i)exp(2\pi i\omega)(sinh\omega\theta + cosh\omega\theta - exp(i\omega(\theta-2\pi)))}{2(cosh\omega\theta + sinh\omega\theta - cos\omega(\theta-2\pi))}\\
\\
(1/b)_\omega=\frac{(1-i)exp(-2\pi i\omega)(sinh\omega\theta + cosh\omega\theta - exp(-i\omega(\theta-2\pi)))}{2(sinh\omega\theta - cos\omega(\theta-2\pi) + cosh\omega\theta)}\\
\end{array}
\end{equation}
$(a/b)_\omega$ is real while $(c/b)_\omega$ and $(1/b)_\omega$ satisfy the following relation
\begin{equation}
(c/b)^{\star}_{\omega} + (1/b)^{\star}_{-\omega} = (c/b)_{-\omega} + (1/b)_\omega
\end{equation}
where $\star$ denotes the complex conjugation.
If $\omega = 0$ then we have
\begin{equation}
\begin{array}{ll}
(a/b)_{\omega = 0} + 1 = 2(\theta-\pi)/\theta&\hspace{15mm} (c/b)_{\omega = 0} + (1/b)_{\omega = 0} = 2(\theta-\pi)/\theta
\end{array}
\end{equation}
Now by using of (23) and the reality of $(a/b)_\omega$ we can define, a set of real ``distributional'' 
orthogonal functions on the $tau=const$ hypersurface 
\begin{equation}
\Phi_\omega(\sigma)=\Theta^+\:\Phi_\omega^E + \Theta^-\:\Phi_\omega^L
\end{equation}
where $\Phi_\omega^E =(\phi^E_\omega(\sigma) + \phi^E_{-\omega}(\sigma)$) , 
$\Phi_\omega^L =(\phi^L_\omega(\sigma) + \phi^L_{-\omega}(\sigma)$) are
symmetric under the interchange of $\omega \leftrightarrow -\omega$.
Then the orthogonality condition of the functions $\Phi_\omega$ is given by 
\begin{equation}
\frac{1}{2\pi}\:\int_0^{2\pi}\!\Phi_\omega(\sigma)\:\Phi_{\omega^\prime}(\sigma) \star 1 = (\delta_{\omega , \omega^\prime} + \delta_{\omega , -\omega^\prime}) <\Phi_\omega,\Phi_{\omega^\prime}>
\end{equation}
They also form a set of complete functions over the hypersurface $\tau=const$.
Therefore we can expand any real function $f(\sigma)$ in terms of them
\begin{equation}
f(\sigma) = \sum_{\omega}\beta_\omega\:\Phi_\omega(\sigma)
\end{equation}
where the coefficients of expansion $\beta_\omega$ and $\beta_{-\omega}$ can 
be determined by using the orthogonality of $\Phi_\omega$ and reality of 
$f(\sigma)$. That is, we have
\begin{equation}
\begin{array}{ll}
\vspace{5mm}\beta_0 = (1/{2\pi}<\Phi_0,\Phi_0>) \int_0^{2\pi}\!f(\sigma)\:\Phi_0 \star 1\hspace{10mm}
\beta_\omega=\beta_{-\omega}=(1/{4\pi}<\Phi_\omega,\Phi_\omega>)\int_0^{2\pi}\!f(\sigma)\:\Phi_\omega\:\star1
\end{array}
\end{equation}
with 
\begin{equation}
\begin{array}{ll}
\vspace{5mm}\Phi_0 =4(\theta-\pi)/\theta &\hspace{30mm}<\Phi_0,\Phi_0> = 16(\theta-\pi)^2/\theta^2
\end{array}
\end{equation}
where eq (24) and the relation $(\Theta^+ + \Theta^- = 1)$ have been used.
Therefore the completeness of $\Phi_\omega$ can be written as
\begin{equation}
\frac{\Phi_0^2}{2\pi<\Phi_0,\Phi_o>} + \sum_{\omega \neq 0}\frac{1}{4\pi}\frac{\Phi_\omega(\sigma)\:\Phi_\omega(\sigma^\prime)}{<\Phi_\omega,\Phi_\omega>}=\delta(\sigma-{\sigma^\prime})
\end{equation}
or
\begin{equation}
\delta(\sigma-\sigma^\prime)= \frac{1}{2\pi} + nonzero  \: modes 
\end{equation}

\section{Poisson bracket structure}
The general solution of this model can be given as a superposition of 
its special solutions which are consistent with the junction condition (10), that is
\begin{equation}
\Phi(\sigma,\tau) = \frac{1}{2\pi}\varphi_0 (\sigma,\tau)+ \sum_{\omega\neq0} \frac{i}{\omega}\alpha_\omega exp(-i\omega\tau)\Phi_\omega(\sigma)
\end{equation}
where $\varphi_0(\sigma,\tau)$ is the zero mode, and $\alpha_\omega$ is the 
normal mode; such that the reality of $\Phi(\sigma,\tau)$ imposes 
the following condition on $\alpha_{\omega}$
\begin{equation}
\alpha^\star_\omega = \alpha_{-\omega}
\end{equation}
The momentum conjugate of $\Phi(\sigma,\tau)$ is given by 
\begin{equation}
\Pi(\sigma,\tau)=  \partial_\tau\:\Phi(\sigma,\tau) \sqrt{\mid g \mid}
\end{equation}
or
\begin{equation}
\Pi(\sigma,\tau) =  \sqrt{\mid g \mid}\: \{ \partial_\tau \varphi_0 + \sum_{\omega \neq0} \alpha_\omega\:\Phi_\omega(\sigma) exp(-i \omega \tau) \}
\end{equation}
Now the poisson  bracket is given by
\begin{equation}
\begin{array}{ll}
\{\Phi(\sigma),\Pi(\sigma^\prime)\} =  \delta(\sigma-\sigma^\prime)\\
\\
\{\Phi(\sigma),\Phi(\sigma')\} = \{\Pi(\sigma),\Pi(\sigma')\} = 0
\end{array}
\end{equation}
By substituting the normal mode expansion of $\Phi$, $\Pi$ and the 
expansion of $\delta(\sigma-\sigma^\prime)$ (30) in (36) we obtain 
\begin{equation}
\begin{array}{ll}
\{\varphi_0,\Pi_0\} = 1 \\
\\
\{\alpha_\omega,\alpha_{\omega^\prime}\} = - i\omega \delta_{\omega+\omega^\prime,0}/{4\pi<\Phi_\omega,\Phi_\omega>}
\end{array}
\end{equation}
where $\Pi_0= \sqrt{\mid g \mid}\: \partial_\tau \varphi_0$. Let us define $\tilde{\alpha}_\omega$ as
\begin{equation}
\tilde{\alpha}_\omega = \alpha_\omega \sqrt{4\pi<\Phi_\omega,\Phi_\omega>}
\end{equation}
Hence the relations (37) and (32) can be rewritten as
\begin{equation}
\{\tilde{\alpha}_\omega , \tilde{\alpha}_{\omega^\prime}\} = - i \omega \delta_{\omega+\omega^\prime,0}
\end{equation}
\begin{equation}
\Phi(\sigma,\tau) = \frac{1}{2\pi}\varphi_0 (\sigma,\tau) + \sum_{\omega\neq0} \frac{i}{\omega}\tilde{\alpha}_\omega exp(-i\omega\tau)\tilde{\Phi}_\omega(\sigma)
\end{equation}
respectively, where 
\begin{equation}
\tilde{\Phi}_\omega(\sigma) = \Phi_\omega(\sigma)/\sqrt{4\pi<\Phi_\omega,\Phi_\omega>}
\end{equation}
                       
\section{Infinite conserved charges}
Energy-momentum tensors are conserved in both (ET) and Lorentzian regions, that is to say
\begin{equation}
\begin{array}{ll}
\partial^{\mu}T_{\mu\nu}^E = 0&\hspace{30mm}0<\sigma<\theta\\
\partial^{\mu}T_{\mu\nu}^L = 0&\hspace{30mm}\theta<\sigma<2\pi
\end{array}
\end{equation}
where by using of (6) and (25) we get the following expressions for the components of 
energy-momentum tensors associated to real scalar field $\Phi(\sigma,\tau)$  
\begin{equation}
\begin{array}{ll}
T_{00}^E =  [(\partial_\tau\Phi^E)^2 - (\partial_\sigma\Phi^E)^2] &\hspace{20mm} T_{01}^E = 2\:\partial_\tau\Phi^E\:\partial_\sigma\Phi^E
\\
T_{00}^L =  [(\partial_\tau\Phi^L)^2 + (\partial_\sigma\Phi^L)^2] &\hspace{20mm} T_{01}^L =  2\:\partial_\tau\Phi^L\:\partial_\sigma\Phi^L
\end{array}
\end{equation}
Introduce new coordinates $\sigma_+^E$ and $\sigma_-^E$ in (ET) region 
\begin{equation}
\begin{array}{ll}
\sigma_+^E = \tau + i\:\sigma\\
\sigma_-^E = \tau - i\:\sigma
\end{array}
\end{equation}
and $\sigma_+^L$ and $\sigma_-^L$ in Lorentzian region 
\begin{equation}
\begin{array}{ll}
\sigma_+^L = \tau + \sigma\\
\sigma_-^L = \tau - \sigma
\end{array}
\end{equation}
In terms of these coordinates in (ET) region the conservation law can be written as
\begin{equation}
\begin{array}{ll}
\partial_+^E\:T_{--}^E = 0\\
\\
\partial_-^E\:T_{++}^E = 0
\end{array}
\end{equation}
and similarly in Lorentzian region we have
\begin{equation}
\begin{array}{ll}
\partial_+^L\:T_{--}^L = 0\\
\\
\partial_-^L\:T_{++}^L = 0
\end{array}
\end{equation}
where
\begin{equation}
\begin{array}{ll}
T_{++}^E = (T_{00}^E - iT_{01}^E)/2 \\
\\
T_{--}^E = (T_{00}^E + iT_{01}^E)/2 \\
\\
T_{+-}^E = T_{-+}^E = 0
\end{array}
\end{equation}
and
\begin{equation}
\begin{array}{ll}
T_{++}^L = (T_{00}^L + T_{01}^L)/2 \\
\\
T_{--}^L = (T_{00}^L - T_{01}^L)/2 \\
\\
T_{+-}^L = T_{-+}^L = 0
\end{array}
\end{equation}
Now we can rewrite the equations (46) and (47) as
\begin{equation}
\begin{array}{ll}
\partial_+^E(f^E(-)\:T_{--}^E) = 0\\
\\
\partial_-^E(f^E(+)\:T_{++}^E) = 0
\end{array}
\end{equation}
and
\begin{equation}
\begin{array}{ll}
\partial_+^L(f^L(-)\:T_{--}^L) = 0\\
\\
\partial_-^L(f^L(+)\:T_{++}^L) = 0
\end{array}
\end{equation}
respectively, where $f(\pm)^E$ and $f(\pm)^L$ are arbitrary functions of $\sigma_\pm^E$ and 
$\sigma_\pm^L$, respectively. So we can get infinite number of conserved quantities. 
Adding the equations (50) and (51) in both regions and writing the partial 
derivatives $\partial_\pm$ in terms of $\partial_\tau$ and $\partial_\sigma$, we get the 
following current conservations
\begin{equation}
\partial_\tau[(f^E(-) + f^E(+)) T_{00}^E + i(f^E(-) - f^E(+))T_{01}^E] + \partial_\sigma[i(f^E(+) - f^E(-))T_{00}^E + (f^E(-) + f^E(+))T_{01}^E] = 0
\end{equation}
in (ET) region, and
\begin{equation}
\partial_\tau[(f^L(-) + f^L(+)) T_{00}^L - (f^L(-) - f^L(+))T_{01}^L] + \partial_\sigma[(f^L(-) - f^L(+))T_{00}^L - (f^L(-) + f^L(+))T_{01}^L] = 0
\end{equation}
in Lorentzian region, respectively.
As in section ${\bf 2}$ in order to have a conserved current across the hypersurface 
of signature change, $\Sigma$, one finds
\begin{equation}
(f^E(-) + f^E(+))\:T_{01}^E\mid_{\Sigma} = - (f^L(-) + f^L(+))\:T_{01}^L\mid_{\Sigma}   
\end{equation}
where, because of similar condition at both hypersurfaces $\Sigma,\Sigma'$, only 
the junction condition at $\Sigma$ is considered. \\
Now, using the junction condition (10) imposed on the $T_{01}$ components in (43), leads 
to the following junction condition over $f$,s 
\begin{equation}
(f^E(-) + f^E(+))\mid_{\Sigma} = (f^L(-) + f^L(+))\mid_{\Sigma}  
\end{equation}
It is quite easy to show that
\begin{equation}
\begin{array}{ll}
\tilde{f}_\omega^E(\sigma^\pm) = [(a/b)_\omega +1] exp(- i \omega{\sigma^E_\pm})/\sqrt{4\pi<\Phi_\omega,\Phi_\omega>}\\
\\
\tilde{f}_\omega^L(\sigma^\pm) = [(c/b)_\omega + (1/b)_{-\omega}]exp(- i\omega{\sigma^L_\pm})/\sqrt{4\pi<\Phi_\omega,\Phi_\omega>}
\end{array}
\end{equation}
and
\begin{equation}
\begin{array}{ll}
\tilde{f}_0^E = \: [(a/b)_0 +1] / \sqrt{<\Phi_0,\Phi_0>}\\
\\
\tilde{f}_0^L = \: [(c/b)_0 + (1/b)_0]/ \sqrt{<\Phi_0,\Phi_0>}
\end{array}
\end{equation}
satisfy the condition (55).
Now one can define the following quantities 
\begin{equation}
\tilde{L}_\omega = \int_0^{2\pi}\!\{\Theta^+[\tilde{f}_\omega^E(+) T_{++}^E +\tilde{ f}_{-\omega}^E(-) T_{--}^E] + \Theta^-[\tilde{f}_\omega^L(+) T_{++}^L + \tilde{f}_{-\omega}^L(-) T_{--}^L]\} d\sigma
\end{equation}
as the required conserved quantities, that is
\begin{equation}
\frac{d \tilde{L}_\omega}{d \tau} = 0
\end{equation}
One can also show that:
\begin{equation}
H = 2 \tilde{L}_0
\end{equation}
where the Hamiltonian is defined as:
\begin{equation}
H = \int_0^{2\pi}\!T_{00}\:\star 1 = \int_0^\theta\!T_{00}^E d\sigma + \int_\theta^{2\pi}\!T_{00}^L d\sigma
\end{equation}

\section{Diffeomorphism algebra}
It is important to note that the conserved quantities $L_\omega$ are closed under 
the Poisson bracket algebra. Considering the following poisson brackets
\begin{equation}
\begin{array}{ll}
\{T_{++}^E(\sigma),T_{++}^E(\sigma^\prime)\} =  i/2\:(T_{++}^E(\sigma) + T_{++}^E(\sigma^\prime))\:\partial_\sigma \:\delta(\sigma-\sigma^\prime)\\
\\
\{T_{--}^E(\sigma),T_{--}^E(\sigma^\prime)\} = - i/2\:(T_{--}^E(\sigma) + T_{--}^E(\sigma^\prime))\:\partial_\sigma \:\delta(\sigma-\sigma^\prime)\\
\\
\{T_{++}^L(\sigma),T_{++}^L(\sigma^\prime)\} = - 1/2\:(T_{++}^L(\sigma) + T_{++}^L(\sigma^\prime))\:\partial_\sigma \:\delta(\sigma-\sigma^\prime)\\
\\
\{T_{--}^L(\sigma),T_{--}^L(\sigma^\prime)\} =  1/2\:(T_{--}^L(\sigma) + T_{--}^L(\sigma^\prime))\:\partial_\sigma \:\delta(\sigma-\sigma^\prime)\\
\\
\{T_{\pm\pm}^E(\sigma),T_{\pm\pm}^L(\sigma')\} = \{T_{\pm\pm}^E(\sigma),T_{\mp\mp}^L(\sigma')\} = 0
\end{array}
\end{equation}
where (48),(49),(43) and (36) has been used, and redefining
\begin{equation}
\tilde{L}_\omega = L(\tilde{f}_\omega^E(+),\tilde{f}_{-\omega}^E(-),\tilde{f}_{\omega}^L(+),\tilde{f}_{-\omega}^L(-))
\end{equation}
the Poisson bracket of $\tilde{L}_\omega$,s takes the following form after some manipulation
\begin{equation}
\{\tilde{L}_\omega,\tilde{L}_{\omega^\prime}\} = -i(\omega - \omega^\prime) L(\tilde{f}_\omega^E(+)\:\tilde{f}_{\omega^\prime}^E(+)\: ,\: \tilde{f}_{-\omega}^E(-)\:\tilde{f}_{-\omega^\prime}^E(-)\: ,\: \tilde{f}_{\omega}^L(+)\:\tilde{f}_{\omega^\prime}^L(+)\: ,\: \tilde{f}_{-\omega}^L(-)\:\tilde{f}_{-\omega^\prime}^L(-))
\end{equation}
It is easy to see that $\tilde{f}_\omega^E(+)$ and  $\tilde{f}_\omega^E(+)\:\tilde{f}_{\omega^\prime}^E(+)$ 
satisfy the equations (13-a,b) and also $\tilde{f}_\omega^L(+)$ and $\tilde{f}_\omega^L(+)\:\tilde{f}_{\omega^\prime}^L(+)$ 
satisfy the equations (15-a,b), therefore one can expand $\tilde{f}_\omega^E(+)\:\tilde{f}_{\omega^\prime}^E(+)$ 
and $\tilde{f}_\omega^L(+)\:\tilde{f}_{\omega^\prime}^L(+)$ in terms of $\tilde{f}_{\omega^{\prime\prime}}^E(+)$ and $\tilde{f}_{\omega^{\prime\prime}}^L(+)$ respectively (2,3-A). 
Hence we can write $\{\tilde{L}_\omega,\tilde{L}_{\omega^\prime}\}$ in terms of 
$\tilde{L}_{\omega^{\prime\prime}}$; or $\tilde{L}_\omega$,s are closed under Poisson bracket 
\begin{equation}
\{\tilde{L}_\omega,\tilde{L}_{\omega^\prime}\} = - i (\omega - \omega^\prime) \sum_{\omega^{\prime\prime}} O_{\omega\omega^\prime}^{\omega^{\prime\prime}}\:\tilde{L}_{\omega^{\prime\prime}}
\end{equation}
where, the structure constants $O_{\omega\omega^\prime}^{\omega^{\prime\prime}}$ 
are real and given by (see appendix A)
\begin{equation}
O_{\omega\omega'}^{\omega''} = \frac{\omega + \omega' + \omega''}{4\pi\omega''[<\tilde{\phi}_{\omega''},\tilde{\phi}_{\omega''}> + <\tilde{\phi}_{\omega''},\tilde{\phi}_{-\omega''}>]}\{\int_0^{\theta}\!\tilde{\phi}_{\omega''}^E(\sigma)\:( \tilde{f}_\omega^E(\sigma)\:
\tilde{f}_{\omega'}^E(\sigma) + \tilde{f}_{-\omega}^E(\sigma)\:\tilde{f}_{-\omega'}^E(\sigma)) d\sigma
\end{equation}
$$
\hspace{65mm} + \int_{\theta}^{2\pi}\! \tilde{\phi}_{\omega''}^L(\sigma)( \tilde{f}_\omega^L(+)\:\tilde{f}_{\omega'}^L(\sigma) + \tilde{f}_{-\omega}^L(\sigma)\:\tilde{f}_{-\omega'}^L(\sigma)) d\sigma\}
$$ 
for $\omega''\neq 0$ and:
\begin{equation}
O_{\omega\omega'}^0 = \frac{1}{4\pi} 
\{\int_0^{\theta}\!(\tilde{f}_\omega^E(\sigma)\:\tilde{f}_{\omega'}^E(\sigma) + \tilde{f}_{-\omega}^E(\sigma)\:\tilde{f}_{-\omega'}^E(\sigma)) d\sigma + \int_{\theta}^{2\pi}\!(\tilde{f}_\omega^L(\sigma)\:\tilde{f}_{\omega'}^L(\sigma) + \tilde{f}_{-\omega}^L(\sigma)\:\tilde{f}_{-\omega'}^L(\sigma) ) d\sigma\}
\end{equation}
for $\omega''= 0$, where $\tilde{\phi}_{\omega}$ are normalized similar to $\tilde{f}_\omega$ by dividing by $\sqrt{4 \pi<\Phi_\omega,\Phi_\omega>}$ 
and $\tilde{f}(\sigma)$,s are $\sigma$ dependent terms in (56).
These structure constants have asymptotic forms (15,19-A) for higher values of $\omega$ 
or $\omega'$. So, by redefinition of $\tilde{L}_\omega$ and $\tilde{L}_0$ as 
\begin{equation}
L_\omega = \frac{1}{\sqrt{|\omega|}} \tilde{L}_\omega , \hspace{20mm} L_0 = \tilde{L}_0
\end{equation}
the poisson bracket (65) is rewritten as 
\begin{equation}
\{L_\omega,L_{\omega^\prime}\} = - i (\omega - \omega^\prime) \sum_{\omega^{\prime\prime}} C_{\omega\omega^\prime}^{\omega^{\prime\prime}}\:L_{\omega^{\prime\prime}}
\end{equation}
where the redefined real structure constants $C_{\omega\omega'}^{\omega''} = O_{\omega\omega'}^{\omega''} \sqrt{|\frac{\omega''}{\omega \omega'}|}$ and 
$C_{\omega\omega'}^0 = O_{\omega\omega'}^0 \frac{1}{\sqrt{|\omega \omega'|}}$ have the 
following asymptotic forms, using of (15,19-A), for higher values of $\omega$ or $\omega'$
\begin{equation}
\begin{array}{ll}
C_{\omega\omega'}^0 \approx \delta_{\omega+\omega'}^0 &\hspace{15mm} 
C_{\omega\omega'}^{\omega''} \approx \delta_{\omega+\omega'}^{\omega''}
\end{array}
\end{equation}
which are the structure constants of ``virasoro'' algebra. 
In table.1 and table.2 a set of numerical structure constants are given for 
two generic values of $\theta$, which confirm (70). Note that for smaller 
$\theta$ the coincidence of structure constants with asymptotic forms are 
more clear than that of larger $\theta$. 
Interpreatation of these asymptotic behaviours will be given in section ${\bf 7}$.

\section{Mode expansion of $L_\omega$ }
Similar to the ordinary 2-dimensional massless boson model, we can expand $L_\omega$ 
in terms of normal modes $\tilde{\alpha}_\omega$.
Hence, using expansion of $\Phi(\sigma,\tau)$, given in (32), we can expand energy 
momentum tensors as
\begin{equation}
\begin{array}{l}
T_{++}^E = \frac{1}{2}[(\partial_\tau\Phi - i \partial_\sigma \Phi)]^2 = 2\:\sum_{\omega\omega'} \tilde{f}_\omega^E(+)\:\tilde{f}_{\omega'}^E(+) \:\tilde{\alpha}_\omega\:\tilde{\alpha}_{\omega'}\\
\\
T_{--}^E = \frac{1}{2}[(\partial_\tau\Phi + i \partial_\sigma \Phi)]^2 = 2\:\sum_{\omega \omega'} \tilde{f}_{-\omega}^E(-)\:\tilde{f}_{-\omega'}^E(-) \:\tilde{\alpha}_\omega\:\tilde{\alpha}_{\omega'}\\       
\\
T_{++}^L = \frac{1}{2}[(\partial_\tau\Phi +  \partial_\sigma \Phi)]^2 = 2\:\sum_{\omega\omega'} \tilde{f}_\omega^L(+)\:\tilde{f}_{\omega'}^L(+) \:\tilde{\alpha}_\omega\:\tilde{\alpha}_{\omega'}\\          
\\
T_{--}^L = \frac{1}{2}[(\partial_\tau\Phi -  \partial_\sigma \Phi)]^2 = 2\:\sum_{\omega\omega'} \tilde{f}_{-\omega}^L(-)\:\tilde{f}_{-\omega'}^L(-) \:\tilde{\alpha}_\omega\:\tilde{\alpha}_{\omega'}          
\end{array}
\end{equation}
Inserting these expansions in formula (58), and using the relations (2,3 - A) for the products 
of $\tilde{f}_\omega$, and also the orthogonality of $\tilde{\Phi}_0$ and $\tilde{\Phi}_\omega$, we 
get the following expansion for $L_\omega$,s
\begin{equation}
L_\omega = 8\pi \sum_{\omega' \omega'' \omega''' } \sqrt{|\omega'\omega''|} C_{\omega' \omega''}^{\omega'''}\:C_{\omega \omega'''}^0 \:\tilde{\alpha}_{\omega'}\: \tilde{\alpha}_{\omega''}
\end{equation}
$L_\omega$ takes the following form for large values of $\omega$
\begin{equation}
L_\omega =8\pi \sum_{\omega' \omega''}\sqrt{|\omega' \omega''|}C_{\omega' \omega''}^{-\omega}\: \tilde{\alpha}_{\omega'}\: \tilde{\alpha}_{\omega''}
\end{equation}
Using the reality of structure constants $C_{\omega \omega'}^{\omega''}$ 
and the relation $\tilde{\alpha}_\omega^\star = \tilde{\alpha}_{-\omega}$ it is easy 
to show that $L_\omega$,s satisfy the following relations
\begin{equation}
L_\omega^\star = L_{-\omega}
\end{equation}
Also the Hamiltonian $H$ has the expansion
\begin{equation}
H = 16\pi \sum_{\omega' \omega''} \sqrt{|\omega'\omega''|} C_{\omega' \omega''}^0\:\tilde{\alpha}_{\omega'} \tilde{\alpha}_{\omega''}
\end{equation}
where we have used eq (70) as
\begin{equation}
C_{0\omega'}^0 \approx \delta_{0\omega'}
\end{equation}
Due to the asymptotic forms of $C_{\omega' \omega''}^0$ and $C_{\omega'\omega''}^{-\omega}$ 
given in eq (70), the higher frequency terms of expansion of $H$ and $L_\omega$ (eqs (75),(73)) will 
look similar to the ordinary two dimensional boson model.
Therfore our model is almost similar to the ordinary boson model for higher modes. 
Physically this is due to the fact that the signature change effect in this model 
is similar to the presentetion of an (ET) wall in the Lorentzian background \cite{A}; so that 
these modifications are reminiscent of a tunneling effect related to the (ET) 
region, in which the rate of tunneling increases for higher energies. Notice that 
due to the existence of a sharp discontinuity in this model it is not expected to 
get the standard results, corresponding to a pure Lorentzian signature, at the 
continuous limit of $\theta \rightarrow 0$; rather, it is seen from quantization 
condition (11) that for small values of $\theta$ the ``sum over energies'' corresponding 
to the spectrum $\omega$ tends to ``sum over integers'' corresponding to the 
ordinary spectrum of boson field, at higher levels of energies (not necessarily 
higher values of energies) compared to the spectrum corresponding to a large value 
of $\theta$. This is because, as $\theta$ tends to smaller values, the cosine function in (11)
oscillates more rapidly and the spectrum is more condensed.

\section{Quantization }

We quantize this model by Dirac canonical quantization. Using the prescription of
\begin{equation}
\{\:\:,\:\:\}_{P.B} \rightarrow \frac{1}{i} [\:\:,\:\:]
\end{equation}
where $[,]$ denotes the commutator, we get the following poisson brackets for operators $\Phi(\sigma),\Pi(\sigma)$ and $\tilde{\alpha}_\omega$
\begin{equation}
\begin{array}{l}
[{\Phi}(\sigma),{\Pi}(\sigma')] = i \delta(\sigma - \sigma')\\

[\Phi(\sigma),\Phi(\sigma')] = [\Pi(\sigma),\Pi(\sigma')] = 0\\

[\tilde{\alpha}_\omega,\tilde{\alpha}_{\omega'}] = \omega \delta_{\omega+\omega',0}
\end{array}
\end{equation}
In order to avoid infinity appearing in operators $L_\omega$ and $H$, 
we have to define them in normal order, similar to the ordinary boson model. Hence, first we define normal ordering for $\tilde{\alpha}_\omega$ as
\begin{equation}
:\tilde{\alpha}_{\omega_1} \: \tilde{\alpha}_{\omega_2}: \doteq \left \{ \begin{array}{ll}
\tilde{\alpha}_{\omega_1}\:\tilde{\alpha}_{\omega_2}&\hspace{15mm}if \:\omega_1<\omega_2\\
\\
\tilde{\alpha}_{\omega_2}\:\tilde{\alpha}_{\omega_1}&\hspace{15mm}if \:\omega_2<\omega_1
\end{array} \right.
\end{equation}
so normal ordered $L_\omega$ is
\begin{equation}
L_\omega = \sum _{\omega'''} [ \sum_{\omega'\leq \omega''}\sqrt{|\omega' \omega''|} 
C_{\omega' \omega''}^{\omega'''}\:C_{\omega \omega'''}^0 \:\tilde{\alpha}_{\omega'}\:\tilde{\alpha}_{\omega''} +  \sum_{\omega''< \omega'} \sqrt{|\omega' \omega''|} C_{\omega' \omega''}^{\omega'''}\:C_{\omega \omega'''}^0 \:\tilde{\alpha}_{\omega''} \:\tilde{\alpha}_{\omega'}]
\end{equation}
Obviously the normal order operator leads to the appearence of a central term or 
anomaly in the diffeomorphism algebra, i.e eq (69)
\begin{equation}
[L_\omega,L_{\omega^\prime}] =  (\omega - \omega^\prime) \sum_{\omega^{\prime\prime}} C_{\omega\omega^\prime}^{\omega^{\prime\prime}}\:L_{\omega^{\prime\prime}} + C(\omega,\omega')
\end{equation}
where the central term $C(\omega,\omega')$ is a function of $\omega$ and $\omega'$.
Here we calculate $C(\omega,\omega')$ for higher values of $\omega$ and $\omega'$. Using eq (73) and
asymptotic behaviour of $C_{\omega \omega'}^{\omega''}$ (70) for higher values of
$\omega,\omega'$ and $\omega''$, $L_\omega$ takes the following form for $\omega > N$
\begin{equation}
L_\omega = \sum_{\omega',\omega''= -N+1}^N \sqrt{|\omega' \omega''|} 
C_{\omega',\omega''}^{-\omega} \: : \tilde{\alpha}_{\omega'}\:\tilde{\alpha}_{\omega''} : + \sum_{\omega'= -\infty}^{-N} \sqrt{|\omega'(\omega + \omega')|} : \tilde{\alpha}_{\omega'} \: \tilde{\alpha}_{-\omega - \omega'} :   \\ 
\end{equation}
$$
\hspace{70mm} + \sum_{\omega'= N+1}^{\infty} \sqrt{|\omega'(\omega+\omega')|}
:\tilde{\alpha}_{\omega'} \: \tilde{\alpha}_{-\omega - \omega'} :
$$
Here $N$ is large enough so that we can take the asymptotic form of $C_{\omega \omega'}^{\omega''}$ (70)
for $\omega,\omega'$ and $\omega''$ larger than $N$. Notice that $N$ is the root of (11) after which 
the roots approaches the roots of $cos{\omega(2\pi- \theta)}$; such that ``sum over energies'' approach to 
``sum over integers''.
Thus for $\omega_1<N$ and $\omega>N$ we get 
\begin{equation}
[\tilde{\alpha}_{\omega_1},L_\omega] = 2 \sum_{\omega'=-N+1}^N \sqrt{|\omega_1 \omega'|} \omega_1\: C_{-\omega_1,\omega'}^{-\omega}\: \tilde{\alpha}_{\omega'} + \sqrt{|\omega_1(\omega-\omega_1)|}\omega_1 \:\tilde{\alpha}_{\omega_1 -\omega}
\end{equation}
and for $\omega_1>N$ and $\omega>N$ we have
\begin{equation}
[\tilde{\alpha}_{\omega_1},L_\omega] = \{\Theta(\omega)[3\Theta(\omega_1)+2\Theta(-\omega_1)] + \Theta(-\omega)[2\theta(\omega_1) + 3\theta(-\omega_1)]\}\omega_1 \sqrt{|\omega_1(\omega-\omega_1)|} \:\tilde{\alpha}_{\omega_1-\omega}
\end{equation}
where $\Theta$ is the step function. Finally, after a rather lenghty calculation 
(for simplicity we assume $\omega,\omega' >0$) and using the asymptotic 
form of $L_\omega$ for higher values of $\omega$, together with the relations 
(83), (84) and (79) the diffeomorphism algebra given in (81) takes the
following form for higher $\omega$ and $\omega'$
\begin{equation}
[L_\omega,L_{\omega^\prime}] =  (\omega - \omega') L_{\omega+\omega'} + C(\omega,\omega')
\end{equation}
with
\begin{equation}
C(\omega,\omega') = \delta_{{\omega+\omega'},0} f(\omega , \omega' ,\theta) -4\sum_{\omega_1 ,\omega_2 >0 }^N {\omega_1}^2 {\omega_2}^2 C_{\omega_1,\omega_2}^{-\omega'}\: C_{\omega_1 ,\omega_2}^{\omega} + \sum_{\omega_1>0}^N {\omega_1}^2\:(\omega - \omega_1)^2\:C_{\omega_1,\omega-\omega_1}^{-\omega'}
\end{equation}
where
$$
f(\omega,\omega',\theta)=3\sum_{l=-n}^0 [-2(l+k)a-\omega-\omega'][2(l+k)a+\omega]^2[|(2(l+k)a+\omega+\omega')(2(l+k)a)|]^{1/2}
$$
and
$$
N=(2k-1)a , \hspace{5mm}  \omega=(2n-1)a, \hspace{5mm}  a=\frac{\pi}{2(2\pi-\theta)}
$$
with $k$ and $n$ as integers.
 
\section{Conclusion }
We have considered a two dimensional model in which space-time is a cylinder 
(circle $\times$ real number) where the circle represents ``space'' and the 
real line represents ``time''. A segment of the circle is fixed with an angle 
$\theta$ and it is assumed that, inside the corresponding infinitely long 
strip-shaped region of the cylinder, the metric is of Euclidean type (ET) 
signature. On one hand, such situations are unrealistic: what happens, when 
an observer tries to enter the (ET) region?
On the other hand, the size of (ET) region might be very small (planck scale?), 
and quantum effects might kill (macroscopic) inconsistencies. 
Nevertheless it is mathematically interesting to investigate which kinds of 
effects a signature changing metric of this type would produce. 
Hence it is worth exploring this model. For the sake of definiteness one may 
put a simple field on this manifold. We have chosen a free massless scalar 
field and have obtained the result that the energy spectrum, diffeomorphism 
algebra and the central term are modified as compared to the standard case (pure Lorentzian signature). 
At least at formal level, these modifications are reminiscent of a tunneling effect related to 
the (ET) region \cite{A}. The modifications of spectrum and diffeomorphism algebra 
tend to disappear for higher values of energies but not the central term.
So it appears that the presentation of signature change affects drastically 
the ordinary boson model at quantum level.
\newpage
{\Large{\ Appendix A }}\\

In order to obtain the structure constants $O_{\omega \omega'}^{\omega''}$ 
given in (65) we note that the left hand side of (65) is time independent. 
Therefore, the right hand side must also be time independent. 
This is possible only if a condition be imposed on the product $ff$, analogous 
to the condition (55) on $f$, as follows
$$
\hspace{23mm}\tilde{f}_\omega^E(+)\:\tilde{f}_{\omega'}^E(+) + \tilde{f}_{-\omega}^E(-)\:\tilde{f}_{-\omega'}^E(-)\mid_{\Sigma} = \tilde{f}_\omega^L(+)\:\tilde{f}_{\omega'}^L(+) + \tilde{f}_{-\omega}^L(-)\:\tilde{f}_{-\omega'}^L(-)\mid_{\Sigma}  \hspace{5mm}(A - 1)
$$
Since $O_{\omega\omega'}^{\omega''}$,s are time independent, then we can find them at the hypersurface
$\tau=const$. 
So we take the following expansions for $\tilde{f}_\omega^E(\sigma)\:\tilde{f}_{\omega'}^E(\sigma)$ 
and $\tilde{f}_\omega^L(\sigma)\:\tilde{f}_{\omega'}^L(\sigma)$ respectively
$$
\hspace{23mm}\tilde{f}_\omega^E(\sigma)\:\tilde{f}_{\omega'}^E(\sigma) = \sum_{\omega''} O_{\omega\omega'}^{\omega''}\:\tilde{f}_{\omega''}^E(\sigma)\hspace{75mm}(A - 2)
$$
$$
\hspace{23mm}\tilde{f}_\omega^L(\sigma)\:\tilde{f}_{\omega'}^L(\sigma) = \sum_{\omega''} O_{\omega\omega'}^{\omega''}\:\tilde{f}_{\omega''}^L(\sigma)\hspace{75mm}(A - 3)                  
$$
where the coefficients $O_{\omega\omega'}^{\omega''}$ are assumed to be the same in 
both (ET) and Lorentzian regions and the functions $\tilde{f}(\sigma)$,s are 
$\sigma$ dependent terms in (56). Since $\tilde{f}_\omega^E(\sigma)$ are real and 
$\tilde{f}_\omega^L(\sigma)$ are complex, we have 
$$
\hspace{23mm}{O_{\omega\omega'}^{\omega''}}^\star = O_{\omega\omega'}^{\omega''} = O_{-\omega -\omega'}^{-\omega''} \hspace{87mm}(A - 4)
$$
it is easily seen from eqs (20) and (56) that
$$
\hspace{23mm}\tilde{f}_\omega^E(\sigma) + \tilde{f}_{-\omega}^E(\sigma) = \: (\tilde{\phi}_\omega^E(\sigma) + \tilde{\phi}_{-\omega}^E(\sigma))\hspace{68mm}(A - 5)
$$
$$ 
\hspace{23mm}\tilde{f}_\omega^L(\sigma) + \tilde{f}_{-\omega}^L(\sigma) = \: (\tilde{\phi}_\omega^L(\sigma) + \tilde{\phi}_{-\omega}^L(\sigma))\hspace{68mm}(A - 6)
$$
with some algebra and with the help of (2-6-A), we get
$$
\hspace{23mm}\tilde{f}_\omega^E(\sigma)\:\tilde{f}_{\omega'}^E(\sigma) + \tilde{f}_{-\omega}^E(\sigma)\:\tilde{f}_{-\omega'}^E(\sigma) =  \sum_{\omega''}(O_{\omega\omega'}^{\omega''} + O_{-\omega -\omega'}^{\omega''})\:\tilde{\phi}_{\omega''}^E(\sigma)\hspace{28mm}(A - 7)
$$
$$
\hspace{23mm}\tilde{f}_\omega^L(\sigma)\:\tilde{f}_{\omega'}^L(\sigma) + \tilde{f}_{-\omega}^L(\sigma)\:\tilde{f}_{-\omega'}^L(\sigma) =  \sum_{\omega''}(O_{\omega\omega'}^{\omega''} + O_{-\omega -\omega'}^{\omega''})\:\tilde{\phi}_{\omega''}^L(\sigma)\hspace{28mm}(A - 8)
$$
Now, we multiply both sides of (7 - A) and (8 - A) by $\tilde{\phi}_{\omega'''}^E(\sigma)$ 
and $\tilde{\phi}_{\omega'''}^L(\sigma)$ and integrate both sides  
from 0 to $\theta$ and $\theta$ to $2\pi$, respectively. By adding these 
relations, and using the orthogonality of $\tilde{\phi}_\omega(\sigma)$,s we get
$$
O_{\omega \omega'}^{\omega''} + O_{-\omega -\omega'}^{\omega''} = \frac{1}{2\pi[<\tilde{\phi}_{\omega''},\tilde{\phi}_{\omega''}> + <\tilde{\phi}_{\omega''},\tilde{\phi}_{-\omega''}>] }\{\int_0^{\theta}\!\tilde{\phi}_{\omega''}^E(\sigma)                                
\:(\tilde{f}_\omega^E(\sigma)\:\tilde{f}_{\omega'}^E(\sigma) + \tilde{f}_{-\omega}^E(\sigma)\:\tilde{f}_{-\omega'}^E(\sigma)) d\sigma                                                                                                                                               
$$
$$
\hspace{69mm} + \int_{\theta}^{2\pi}\! \tilde{\phi}_{\omega'''}^L(\sigma)(\tilde{f}_\omega^L(\sigma)\:\tilde{f}_{\omega'}^L(\sigma) + \tilde{f}_{-\omega}^L(\sigma)\:\tilde{f}_{-\omega'}^L(\sigma))d\sigma \}\hspace{5mm}(A - 9)
$$
On the other hand, in addition to the junction condition (55), one can find another 
condition over the functions $\tilde{f}_\omega^E(\sigma)$ and 
$\tilde{f}_\omega^L(\sigma)$ (56), namely
$$
\hspace{23mm}\tilde{f}_\omega^E(+) - \tilde{f}_{-\omega}^E(-) \mid_{\Sigma} =i(\tilde{f}_{\omega}^L(+) - \tilde{f}_{-\omega}^L(-)) \mid_{\Sigma}\hspace{55mm}(A - 10)
$$
Using (2, 3 - A) and (5, 6 - A), one can show that 
$$
\hspace{20mm}\tilde{f}_\omega^E(\sigma)\:\tilde{f}_{\omega'}^E(\sigma) - \tilde{f}_{-\omega}^E(\sigma)\:\tilde{f}_{-\omega'}^E(\sigma) =  \sum_{\omega''}\frac{1}{\omega''}(O_{\omega\omega'}^{\omega''} - O_{-\omega -\omega'}^{\omega''})\:\partial_{\sigma}\:\tilde{\phi}_{\omega''}^E(\sigma)\hspace{22mm}(A - 11)
$$
$$
\hspace{20mm}-i\:(\tilde{f}_\omega^L(\sigma)\:\tilde{f}_{\omega'}^L(\sigma) - \tilde{f}_{-\omega}^L(\sigma)\:\tilde{f}_{-\omega'}^L(\sigma)) =  \sum_{\omega''}\frac{1}{\omega''}(O_{\omega\omega'}^{\omega''} - O_{-\omega -\omega'}^{\omega''})\:\partial_{\sigma}\:\tilde{\phi}_{\omega''}^L(\sigma)\hspace{15mm}(A - 12)
$$
Now, we multiply both sides of (11 - A) and (12 - A) by 
$\partial_{\sigma}\:\tilde{\phi}_{\omega'''}^E(\sigma)$ and 
$\partial_{\sigma}\:\tilde{\phi}_{\omega'''}^L(\sigma)$ respectively and
integrate in the same way as before. Then, subtracting the two integrated 
expressions, the surface terms cancel each other by using (10 - A) and 
(1 - A). Using the orthogonality of $\tilde{\phi}_\omega(\sigma)$,s we get
$$
O_{\omega\omega'}^{\omega''} - O_{-\omega -\omega'}^{\omega''} = 
\frac{\omega + \omega'}{2\pi\omega''[<\tilde{\phi}_{\omega''},\tilde{\phi}_{\omega''}> + <\tilde{\phi}_{\omega''},\tilde{\phi}_{-\omega''}>]}\{\int_0^{\theta}\!\tilde{\phi}_{\omega''}^E(\sigma)\:(\tilde{f}_\omega^E(\sigma)\:\tilde{f}_{\omega'}^E(\sigma) + \tilde{f}_{-\omega}^E(\sigma)\:\tilde{f}_{-\omega'}^E(\sigma))d\sigma  
$$
$$
\hspace{70mm} + \int_{\theta}^{2\pi}\! \tilde{\phi}_{\omega''}^L(\sigma)(\tilde{f}_\omega^L(\sigma)\:\tilde{f}_{\omega'}^L(\sigma) + \tilde{f}_{-\omega}^L(\sigma)\:\tilde{f}_{-\omega'}^L(\sigma))d\sigma\}\hspace{5mm}(A - 13)
$$
Finally, using the relations (9 - A) and (13 - A), we obtain the following 
expression for the coefficients given in (66)
$$
O_{\omega\omega'}^{\omega''} = \frac{\omega + \omega' + \omega''}{4\pi\omega''[<\tilde{\phi}_{\omega''},\tilde{\phi}_{\omega''}> + <\tilde{\phi}_{\omega''},\tilde{\phi}_{-\omega''}>]}\{\int_0^{\theta}\!\tilde{\phi}_{\omega''}^E(\sigma)\:(\tilde{f}_\omega^E(\sigma)\:
\tilde{f}_{\omega'}^E(\sigma) + \tilde{f}_{-\omega}^E(\sigma)\:\tilde{f}_{-\omega'}^E(\sigma)) d\sigma 
$$
$$
\hspace{70mm} + \int_{\theta}^{2\pi}\! \tilde{\phi}_{\omega''}^L(\sigma)(\tilde{f}_\omega^L(\sigma)\:\tilde{f}_{\omega'}^L(\sigma) + \tilde{f}_{-\omega}^L(\sigma)\:\tilde{f}_{-\omega'}^L(\sigma))d\sigma\}\hspace{2mm}(A - 14)
$$
one can show from (14-A) or (2,3-A) together with (56) that the limits of $\omega$, $\omega' \rightarrow \infty$ 
tend to the asymptotic form
$$
\hspace{27mm} O_{\omega\omega'}^{\omega''} \approx \sqrt{\frac{\omega \omega'}{\omega''}} \delta_{\omega + \omega'}^{\omega''}\hspace{90mm}(A - 15)
$$
For $\omega''= 0$, we rewrite (7-A) and (8-A) as
$$
\hspace{23mm}\tilde{f}_\omega^E(\sigma)\:\tilde{f}_{\omega'}^E(\sigma) + \tilde{f}_{-\omega}^E(\sigma)\:\tilde{f}_{-\omega'}^E(\sigma) =  \sum_{\omega''}(O_{\omega\omega'}^{\omega''})(\tilde{\phi}_{\omega''}^E(\sigma) + \tilde{\phi}_{-\omega''}^E(\sigma))\hspace{25mm}(A - 16)   
$$
$$
\hspace{23mm}\tilde{f}_\omega^L(\sigma)\:\tilde{f}_{\omega'}^L(\sigma) + \tilde{f}_{-\omega}^L(\sigma)\:\tilde{f}_{-\omega'}^L(\sigma) =  \sum_{\omega''}(O_{\omega\omega'}^{\omega''})(\tilde{\phi}_{\omega''}^L(\sigma) + \tilde{\phi}_{-\omega''}^L(\sigma))\hspace{25mm}(A - 17)
$$
where (4-A) has been used.
By integrating both sides of (16-A) and (17-A) from 0 to $\theta$ and 
$\theta$ to $2\pi$ respectively, and adding these two relations and using the 
orthogonality of $\tilde{\Phi}_\omega$ and $\tilde{\Phi}_0$ we finally get
$$
O_{\omega\omega'}^0 = \frac{1}{4\pi} \{\int_0^{\theta}\!(\tilde{f}_\omega^E(\sigma)
\:\tilde{f}_{\omega'}^E(\sigma) + \tilde{f}_{-\omega}^E(\sigma)\:\tilde{f}_{-\omega'}^E(\sigma)) d\sigma + \int_{\theta}^{2\pi}\!(\tilde{f}_\omega^L(\sigma)\:\tilde{f}_{\omega'}^L(\sigma) + \tilde{f}_{-\omega}^L(\sigma)\:\tilde{f}_{-\omega'}^L(\sigma)) d\sigma\}\hspace{5mm}(A - 18)
$$
In the limits of $\omega$, $\omega'\rightarrow\infty$, equations (2,3-A) together with (56),(57) 
or (18-A) with the help of the orthogonality of $\tilde{\Phi}_{\omega+\omega'}$ and $\tilde{\Phi}_0$ give the asymptotic form  

$$
\hspace{27mm}O_{\omega\omega'}^0 \approx \delta_{\omega+\omega'}^0 \sqrt{\omega \omega'} \hspace{90mm}(A - 19)
$$
\setcounter{equation}{76}
\vspace{2mm}

{\large{\bf Acknowledgements}}
\vskip8pt\noindent
We wish to thank H.R.Sepangi for creating interest among us concerning the 
subject of signature change. Also we thank S.K.A.Seyed-Yagoobi, A.R.Rastegar and M.H.Feizi-Derakhshi for
reading the manuscript also their constructive comments and discussions.

\newpage
\begin{center}
\hspace{15mm}Table 1 
\hspace{5mm} $\theta$ = 1.0000000000E-02    rad
\begin{tabular}{|c|c|c|c|} \hline
$\omega1$ & $\omega2$ & $\omega3$ & $C_{\omega1,\omega2}^{\omega3}$ \\ \hline
4.0000016970E+00 & 1.0000000265E+00 & 3.0095790648E+00 &  7.0319114695E-12 \\ \hline    
4.0000016970E+00 & 1.0000000265E+00 & 5.0000033137E+00 &  7.0710443953E-01 \\ \hline    
4.0127713346E+00 & 1.0000000265E+00 & 3.0000007161E+00 &  4.0600626371E-10 \\ \hline   
4.0127713346E+00 & 1.0000000265E+00 & 5.0000033137E+00 &  3.9199551970E-08 \\ \hline    
4.0127713346E+00 & 1.0000000265E+00 & 5.0159629605E+00 &  7.0709265443E-01 \\ \hline     
4.0000016970E+00 & 1.0000000265E+00 & 1.0000000265E+00 & -5.6216156909E-08 \\ \hline
4.0000016970E+00 & 1.0000000265E+00 & 4.0127713346E+00 &  1.3576975933E-11 \\ \hline
4.0127713346E+00 & 1.0000000265E+00 & 4.0000016970E+00 &  6.1380770510E-10 \\ \hline
4.0127713346E+00 & 1.0000000265E+00 & 2.0000002122E+00 &  4.2912917240E-10 \\ \hline
4.0000016970E+00 & 1.0000000265E+00 & 1.0031932364E+00 &  1.1307372973E-11 \\ \hline
\end{tabular}
\end{center}
\vspace{20mm}
\begin{center}
\hspace{15mm}Table 2
\hspace{5mm} $\theta$ = 7.8539000000E-01     rad
\begin{tabular}{|c|c|c|c|} \hline
$\omega1$ & $\omega2$ & $\omega3$ & $C_{\omega1,\omega2}^{\omega3}$ \\ \hline
7.1415132092E+00 & 5.9967146145E+00 & 1.2285719502E+01 & -2.3137928725E-01 \\ \hline   
7.1415132092E+00 & 5.9967146145E+00 & 1.2857108790E+01 & -3.8938003617E-10 \\ \hline   
7.1415132092E+00 & 5.9967146145E+00 & 1.3428561050E+01 &  6.7943095384E-01 \\ \hline
7.1415132092E+00 & 5.9967146145E+00 & 1.3999973109E+01 & -2.5636294321E-10 \\ \hline
7.1415132092E+00 & 5.9967146145E+00 & 1.6285689090E+01 & -2.4443330520E-10 \\ \hline
8.2851589177E+00 & 8.2851589177E+00 & 1.2285719502E+01 & -4.5534806962E-02 \\ \hline
8.2851589177E+00 & 8.2851589177E+00 & 1.2857108790E+01 &  1.1662070168E-09 \\ \hline
8.2851589177E+00 & 8.2851589177E+00 & 1.6285689090E+01 &  1.0397107396E-09 \\ \hline
8.2851589177E+00 & 8.2851589177E+00 & 1.6857118474E+01 &  6.8031603221E-01 \\ \hline   
8.2851589177E+00 & 8.2851589177E+00 & 1.7428545137E+01 &  8.0611066518E-10 \\ \hline                         
\end{tabular}
\end{center}
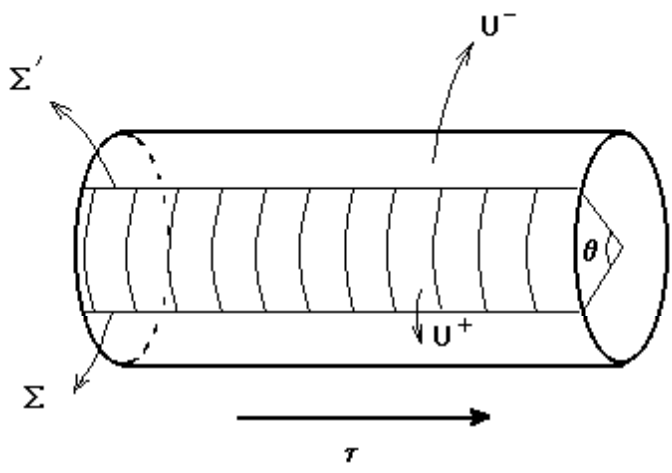
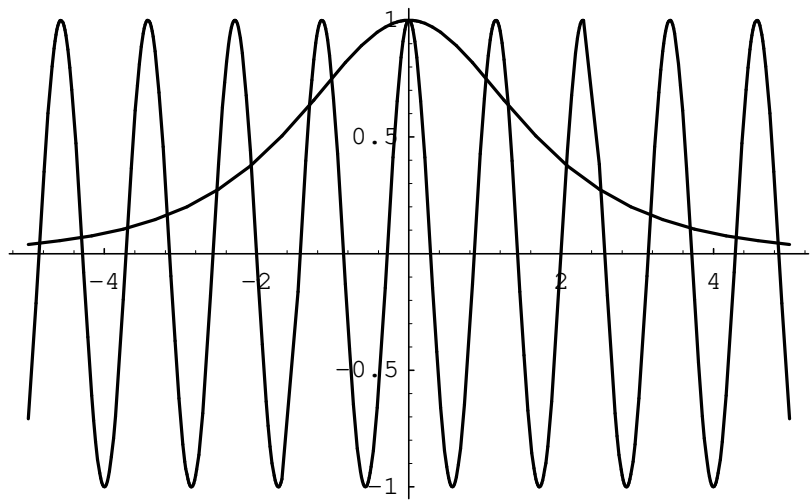
\end{document}